\documentclass[a4paper,10pt]{article}
\usepackage{amsfonts}
\usepackage{amsmath}
\usepackage{amssymb,youngtab}
\usepackage{enumerate}
\usepackage{amsthm}
\usepackage[pdftex]{graphicx}

\title{Laplace maps and constraints for a class of third order partial differential operators }
\author{Chris Athorne}

\begin{document}

\maketitle

\begin{abstract} We explore the existence of a class of generalised Laplace maps for third order partial differential operators of the form
\[\partial_1\partial_2\partial_3+a_1\partial_2\partial_3+a_2\partial_1\partial_3+a_3\partial_1\partial_2+a_{12}\partial_3+a_{23}\partial_1+a_{13}\partial_2+a_{123}\] 
and related first order $3\time 3$ systems and show that they require the satisfaction of constraints on the invariants for such operators.
\end{abstract}

\section{Introduction} The classical Laplace maps are transformations of an essentially algebraic nature between second order, linear, partial differential operators. They were developed by Darboux in \cite{D} and can be thought about in a variety of ways with application to several areas. They are (generically invertible) maps between invariants associated with different operators.

The nomenclature surrounding this type of map is neither fixed nor always clear. So it is that the classical maps are sometimes described as Laplace transformations (rather than maps), as belonging to the more general class of \emph{Darboux transformations} and as \emph{intertwining} Laplace transformations (in order to distinguish this approach from other possible avenues of generlization). Such maps may also be considered to act either on the partial differential operators themselves or on the elements of their kernels, the homogeneous solution spaces.

In the theory of integrable sytems the Toda lattice \cite{T} has been an influential paradigm. It is related to the theory of Laplace maps in that the maps generate three term recurrence relations on the (indexed) invariants which are exactly the equations of the Toda chain on $\mathbb Z$ over $\mathbb R^2$ \cite{W}. The vanishing of an invariant corresponds to the end of the chain and to a factorizable operator. From the kernel of this factorizable case all the invariants on the chain can be generated. 

The Laplace maps can also be lifted to transformations of non-linear systems of hydrodynamic type because the Riemann invariants satisfy second order, linear partial differential equations \cite{F,ZS}.

In geometry the transformations describe maps between immersed surfaces and their accompanying conjugate nets (coordinates), a connection generalised to higher dimensions in \cite{Kam, KamT}. This relation can be developed in discrete geometry also \cite{Dol, N}.

As geometrical objects the invariants themselves can be derived following the Cartan method of moving frames \cite{SM}.

From a purely algebraic point of view, Laplace maps have been studied in work on factorization \cite{BKar,JA,SW1,SW2,Ts1,Ts2}. The general feeling appears to be that the classical instance of Laplace maps is not easily generalizable.

Further classes of generalization include \emph{Darboux transformations of type I} \cite{S1,S2,S3} and of \emph{continued type} \cite{HS}. A looser and more general notion of \emph{intertwining Laplace transformation} than that presented in this paper can be found in \cite{G}.

Generalizations of an even less classical nature involve quasideterminants \cite{LN} and supersymmetry or geometry \cite{LM,LSV,HSV}.

This paper has two immediate predecessors. The first is \cite{A} where Laplace maps for $3\times 3$ systems are discussed. The current paper will slightly generalise that work. The second is \cite{AY} where a classification of certain invariants for a very large class of arbitrary order invariants is given.

Maps of this sort are usually characterised by an intertwining property of the form
\[A^\sigma a=a^\sigma A\]
where one is interested in the kernel of $A^\sigma$ in relation to that of $A$: 
\[a:\ker A\rightarrow\ker A^\sigma.\]

In this account we describe several classes of intertwining relation for third order partial differential operators in scalar and system form and show that intertwining maps of this type can exist only when certain constraints on the invariants of the operators are satisfied. This is perhaps, compared with the second order case, a disappointing result but it is compatible with the general difficulty of generalisation apparent in the literature. It is also the case that these constraints are not preserved as functions of the transformed invariants so that one may not successively apply the Laplace maps.

Further we compare the results for scalar and system forms. It is not generally true that scalar third order partial differential equations can be written in $3\times 3$ system form nor vici versa. Again, certain conditions on invariants have to be satisfied and we look at the way the Laplace maps correspond under these conditions.

In section 2 we review the classical situation, presenting it in a form convenient to generalisation by introducing invariants as zeroth order differential operators constructed from the natural first and second order operators of the theory. 

In the next section a technical trick is introduced which allows us to analyse polynomial identities between differential operators by reducing them to successively lower order identities and we introduce a deformation of the standard Laplace map.

Section 4 attempts to generalise the approach of section 2 to third order differential operators considering two scenarios: firstly first order and then second order intertwining operators. We find that constraints arise in each case, but they are fewer in number for the second order intertwiner. We then review the $3\times 3$ system. Laplace maps for this case have been considered before but we clarify here the role of constraints needed for the Laplace map to function and we discuss the relation between scalar and system forms and their overlap when a Laplace map exists.

Throughout the paper our philosophy is to work with noncommutative polynomials generated by a finite set of (noncommuting) linear differential operators. In particular our starting point is always a single linear operator of total degree $n$ in derivations $\partial_1,\ldots,\partial_n$ but which is of degree one only in each distinct derivation. In \cite{AY} these were called \emph{hyperbolic} but this suggests a reality condition that does not feature in the discussion.

Alhough in that paper systematic processes describe the construction of invariants, the calculations in the current paper are somewhat \emph{ad hoc} and a deeper understanding of their structure would be required for discussion of higher orders to be feasible.

Finally we make some concluding remarks and suggestions for further exploration.

\section{The Classical case}
\subsection{Scalar form}
Consider the second order partial differential operator
\[ L_{12}=\partial_1\partial_2+a_2\partial_1+a_1\partial_2+a_{12}.\]
The coefficients belong to a differential field with derivations $\partial_1$ and $\partial_2$ and we assume no relations (algebraic or differential) between them. We write $a_i,_j$ for the $\partial_j$ derivative of $a_i$ etc. Note that the operator is symmetric in indices: $L_{12}=L_{21}.$

We are interested in invariants of such operators under transformations of the form: $L_{12}\mapsto L_{12}^g=g^{-1}L_{12}g$ where $g$ is an arbitrary element of the differential field or an extension thereof. Invariants are constructed by defining
\[L_1=\partial_1+a_1,\quad L_2=\partial_2+a_2\]
and writing down the functions
\[I_{12}=L_{12}-L_1L_2,\quad I_{21}=L_{12}-L_2L_1.\]
These are invariants by virtue of being differential functions (zeroth order operators) of the coefficients:
\begin{eqnarray}
I^g_{12}&=&L^g_{12}-L^g_1L^g_2\nonumber\\
&=&g^{-1}(L_{12}-L_1L_2)g\nonumber\\
&=&I_{12}\nonumber
\end{eqnarray}

Thus
\[L_{12}=L_1L_2+I_{12}=L_2L_1+I_{21}\]
where $I_{12}=a_{12}-a_1a_2-a_{2,1}$ and $I_{21}=a_{12}-a_1a_2-a_{1,2}.$  

Now suppose there is an element $\phi$ (in a field extension) such that $L_{12}\phi=0.$ Define $\phi^\sigma=L_2\phi$ and $\phi^\Sigma=L_1\phi.$ These satisfy the pairs
\begin{eqnarray}
L_2\phi&=&\phi^\sigma\nonumber\\
L_1\phi^\sigma&=&-I_{12}\phi
\end{eqnarray}
and
\begin{eqnarray}
L_1\phi&=&\phi^\Sigma\nonumber\\
L_2\phi^\Sigma&=&-I_{21}\phi
\end{eqnarray}

The $\sigma-$ and $\Sigma-$Laplace transformed equations are those satisfied by $\phi^\sigma$ and $\phi^\Sigma$ obtained by eliminating $\phi$ from the above pairs.

\[(L_2^\sigma L_1+I_{12})\phi^\sigma=(L_1 L_2^\sigma+I_{12}+[L_2^\sigma,L_1])\phi^\sigma=0\]
\[(L_1^\Sigma L_2+I_{21})\phi^\Sigma=(L_2L_1^\Sigma + I_{21}+[L_1^\Sigma,L_2])\phi^\Sigma=0\]
where $L_2^\sigma=I_{12}L_2I_{12}^{-1}$ and $L_1^\Sigma=I_{21}L_2I_{21}^{-1}$. This implies transformations on the invariants:

\begin{eqnarray}
I_{21}^\sigma&=&I_{12}\nonumber\\
I_{12}^\sigma&=&I_{12}+[L_2^\sigma,L_1]\nonumber\\
I_{21}^\Sigma&=&I_{21}+[L_1^\Sigma,L_2]\nonumber\\
I_{12}^\Sigma&=&I_{21}\nonumber
\end{eqnarray}
and correspondingly:
\[L_1^\sigma=L_1,\quad L_2^\Sigma=L_2.\]

These relations are all summarised in the simple intertwining relations
\[L^\sigma_{12} L_2=L_2^\sigma L_{12}\]
\[L^\Sigma_{12} L_1=L_1^\Sigma L_{12}.\]

For example, the first implies the following chain of argument.
\begin{eqnarray}
(L_2^\sigma L_1^\sigma+I_{21}^\sigma)L_2&=&L_2^\sigma(L_1L_2+I_{12}),\nonumber\\
L_2^\sigma(L_1^\sigma-L_1)L_2&=&L_2^\sigma I_{12}-I^\sigma_{21}L_2.\nonumber
\end{eqnarray}
By looking at leading order terms in differential operators,
\begin{eqnarray}
L_1^\sigma&=&L_1\nonumber\\
I_{21}^\sigma&=&I_{12}\nonumber\\
L_2^\sigma&=&I_{12}L_2I_{12}^{-1}\nonumber
\end{eqnarray}
and finally, since $L^\sigma_{12}=L^\sigma_{21},$
\begin{eqnarray}
I^\sigma_{12}&=&I^\sigma_{21}+[L^\sigma_2,L^\sigma_1]\nonumber\\
&=&I_{12}+[L^\sigma_2,L_1]\nonumber\\
&=&2I_{12}-I_{21}+(\log I_{12}),_{12}.\nonumber
\end{eqnarray}

We can however, also deform the intertwining relation by writing:
\begin{equation}\label{def'2}
L^\sigma_{12}L'_2=L'^\sigma_{2}L_{12},
\end{equation}
where the primed operators are monic in $\partial_2$ still but with coefficients distinct from the unprimed operators. We consider this case at the conclusion of the next section.

\subsection{System form}
The situation just described can always be represented in system form. We can write the pairs of equations for $\phi^\sigma$ and $\phi^\Sigma$ as

\begin{eqnarray}
\left(
\begin{array}{cc}
L_2 & -1 \\
I_{12} & L_1
\end{array}
\right)
\left(
\begin{array}{c}
\phi\\
\phi^\sigma
\end{array}
\right)
&=&
\left(
\begin{array}{c}
0\\
0
\end{array}
\right)
\nonumber\\
\left(
\begin{array}{cc}
L_2 & I_{21} \\
-1 & L_1
\end{array}
\right)
\left(
\begin{array}{c}
\phi^\Sigma\\
\phi
\end{array}
\right)
&=&
\left(
\begin{array}{c}
0\\
0
\end{array}
\right)\nonumber
\end{eqnarray}
and the intertwining relations become
\[
\left(
\begin{array}{cc}
L^\sigma_2 & -1 \\
I_{12}^\sigma & L^\sigma_1
\end{array}
\right)
\left(
\begin{array}{cc}
L_2 & 0 \\
0 & L^\sigma_2
\end{array}
\right)
=
\left(
\begin{array}{cc}
L^\sigma_2 & 0\\
{[}L_2^\sigma, L_1{]} & L_2^\sigma
\end{array}
\right)
\left(
\begin{array}{cc}
L_2 & -1\\
I_{12} & L_1
\end{array}
\right)
\]
and
\[
\left(
\begin{array}{cc}
L^\Sigma_2 & I^\Sigma_{21}\\
-1 & L^\Sigma_1
\end{array}
\right)
\left(
\begin{array}{cc}
L_1^\Sigma & 0 \\
0 & L_1
\end{array}
\right)
=
\left(
\begin{array}{cc}
L^\Sigma_1 & [L_1^\Sigma, L_2]\\
0 & L_1^\Sigma
\end{array}
\right)
\left(
\begin{array}{cc}
L_2 & I_{21}\\
-1 & L_1
\end{array}
\right)
\]

\section{Methodology}
We adopt some ideas from the paper \cite{AY} where invariants of arbitrary order differential operators were constructed.

Let $I$ be a set of distinct labels. We associate with $I$ a partial differential operator in derivations $\partial_i$ for $i\in I:$ 
\[L_I=\sum_{J\subseteq I}a_{I\backslash J}\partial_J.\]
Here the symbol $\partial_J$ is the product of all derivations $\partial_j$ for $j\in J.$ All the coefficients $a_K$ for $K\subseteq I$ are algebraically and differentially independent. We associate such an operator with any subset of $I$ in a similar manner.

The coefficients are totally symmetric in their indices so that $L_I$ is a function on the set; that is, $L_I$ is totally symmetric in its indices. 

For example if $I=\{1,2,3\},$
\begin{eqnarray}
L_{123}&=&\partial_1\partial_2\partial_3+a_1\partial_2\partial_3+a_2\partial_3\partial_1+a_3\partial_1\partial_2\nonumber\\
&&+a_{12}\partial_3+a_{23}\partial_1+a_{31}\partial_2+a_{123}\nonumber\\
L_{12}&=&\partial_1\partial_2+a_1\partial_2+a_2\partial_1+a_{12}\nonumber\\
L_{23}&=&\partial_2\partial_3+a_2\partial_3+a_3\partial_2+a_{23}\nonumber\\
L_{31}&=&\partial_3\partial_1+a_3\partial_1+a_1\partial_3+a_{31}\nonumber\\
L_1&=&\partial_1+a_1\nonumber\\
L_2&=&\partial_2+a_2\nonumber\\
L_3&=&\partial_3+a_3\nonumber
\end{eqnarray}

We consider the non-commutative polynomial ring over in these operators over some constants $K:$
\[K_I=K[\{L_J|J\subseteq I\}].\]

The invariants are the zeroth order differential elements of this ring. For example, in $K_{\{1,2\}}$ both $L_{12}-L_1L_2$ and $L_{21}-L_2L_1$ are zeroth order. Quite generally, for any index set $I,$  because the differential operators transform as $L_J\mapsto g^{-1}L_Jg,$ for an arbitrary function $g,$ any polynomial in the $L_J,$ $F(L_J|J\subseteq I),$ transforms similarly: $F\mapsto g^{-1}Fg.$ In the case where such an $F$ happens to be a zeroth order operator (i.e. a function), it is therefore invariant. 

In the paper \cite{AY} a large class of invariants is constructed.

We can describe such invariant elements by finding the kernel of a map $\Theta$ defined by
\[\Theta(L)=[L,\theta]\]
$\theta$ being regarded as an arbitrary function.

$\Theta$ acts as a derivation on the ring $K_I.$ We can then take a set of such maps $\{\Theta_i|i\in I\}$ corresponding to differentiation of elements of $K_I$ with respect to indices by choosing $\theta=x_i$. 

We illustrate the methodology by a redescription of the classical case. Thus
\[\Theta_1(L_{12})=L_2,\quad \Theta_1(L_2)=0,\quad \Theta_1(L_1)=1,\]
and so on. In particular we see that $I_{12}$ and $I_{21}$ are invariants because
\begin{eqnarray}
\Theta_1(L_{12}-L_1L_2)&=&L_2-L_2=0,\nonumber\\
\Theta_2(L_{12}-L_1L_2)&=&L_1-L_1=0,\nonumber
\end{eqnarray}
and similarly for $I_{21}.$

We can use the $\Theta_i$ maps to analyse the intertwining relation. By applying $\Theta_1$ and $\Theta_2$ successively we obtain, for the $\sigma$ case for example, the following non trivial relations:
\begin{eqnarray}
L^\sigma_{12}L_2&=&L_2^\sigma L_{12}\nonumber\\
L^\sigma_1L_2+L^\sigma_{12}&=&L_{12}+L_2^\sigma L_1\nonumber\\
2L_1^\sigma&=&2L_1\nonumber.
\end{eqnarray}

The last tells us that $L_1^\sigma=L_1$ and we then rearrange the second to give $I^\sigma_{21}=I_{12}.$ The first equation can then be written
\[(L_{12}^\sigma-L_2^\sigma L_1^\sigma)L_2=L^\sigma_2(L_{12}-L_1L_2),\]
equivalently
\[L_2^\sigma=I_{12}L_2I^{-1}_{12},\]
and finally
\[I^\sigma_{12}=I^\sigma_{21}+[L_2^\sigma,L_1^\sigma]=I_{12}+[L_2^\sigma,L_1].\]

Returning to the deformed case (\ref{def'2}) we may analyse this is the same way. By derivation using $\Theta_1$ and $\Theta_2$ we obtain the tower:
\begin{eqnarray}
L^\sigma_{12}L'_2&=&L'^\sigma_2L_{12}\nonumber\\
L^\sigma_2L'_2&=&L'^\sigma_2L_2\nonumber\\
L^\sigma_1L'_2+L^\sigma_{12}&=&L_{12}+L'^\sigma_2L_1\nonumber\\
L^\sigma_2+L'_2&=&L'^\sigma_2+L_2\nonumber\\
L_1^\sigma&=&L_1.\nonumber
\end{eqnarray}

From these equations (in reverse order) follow:
\begin{eqnarray}
L^\sigma_1&=&L_1\nonumber\\
L'^\sigma_2&=&L^\sigma_2+\phi\nonumber\\
L'_2&=&L_2+\phi\nonumber\\
L^\sigma_2\phi&=&\phi L_2\nonumber\\
I^\sigma_{21}&=&I_{12}-\phi,_1\label{int1}\\
I^\sigma_{21}\phi,_2-\phi I^\sigma_{21},_2&=&\phi^2(I_{21}-I^\sigma_{21})\label{int2}
\end{eqnarray}

Note that this defomation is not equivalent to the undeformed case: we cannot gauge away the $\phi$ term in the primed operators by writing $L'_2=g^{-1}L_2g$ without introducing compensating terms in $L_{12}$ and so on.

From equation (\ref{int2}) we obtain, by putting 
\[\phi=\frac{I^\sigma_{21}}{\psi}\]
the relation
\[I^\sigma_{21}=I_{21}+\psi,_2=I_{21}+\left(\frac{I^\sigma_{21}}{\phi}\right),_2.\]

Then $\phi$ is determined by the invariants $I_{12}$ and $I_{21}$ via the following equation:
\[I_{12}-I_{21}=\left(\frac{I_{12}}{\phi}\right),_2+\phi,_1-(\log\phi),_{12}.\]

In the limit that $\phi,$ but not $\phi^{-1}\phi,_2,$ tends to zero, we recover the classical case:
\[\phi^{-1}\phi,_2=I^{-1}_{12}I_{12},_2\]
and $\phi=I_{12}.$

Otherwise this equation looks not easy to solve and is the object of further study.

\section{Third order case}
In the third order case, $I=\{1,2,3\}$ there are six invariants on two labels,
\[I_{ij}=L_{ij}-L_iL_j,\quad i,j\in \{1,2,3\}.\]
The Jacobi identity,
\[[L_1,[L_2,L_3]]+[L_2,[L_3,L_1]]+[L_3,[L_1,L_2]]=0,\]
yields a single identity 
\[I_{12},_3+I_{23},_1+I_{31},_2-I_{21},_3-I_{32},_1-I_{13},_2=0.\]

On three labels we have the set
\[I_{ijk}=L_{ijk}-I_{jk}L_i-I_{ik}L_j-I_{ij}L_k-L_iL_jL_k.\]
Because of the total symmetry of $L_{ijk}$ we may make any of these our choice for one independent invariant. Indeed under transpositions of indices:
\begin{eqnarray}
I_{jik}&=&I_{ijk}\nonumber\\
I_{ikj}&=&I_{ijk}+(I_{kj}-I_{jk}),_i\nonumber\\
I_{kji}&=&I_{ijk}+(I_{ki}-I_{ik}),_j\nonumber.
\end{eqnarray}

One checks the invariance of $I_{123}$ by, for example,
\[\Theta_1(I_{123})=L_{23}-I_{23}-L_2L_3=0\]
and likewise with $\Theta_2$ and $\Theta_3.$

We call the expression of $L_{ijk}$ in terms of invariants and the $L_i,$
\[L_{ijk}=L_iL_jL_k+I_{jk}L_i+I_{ik}L_j+I_{ij}L_k+I_{ijk},\]
the {\em invariant expansion} of $L_{ijk}$.

We will discuss the system form in a moment. However, it should be emphasised that scalar and system forms are not interchangeable at order three unlike the order two case. We are generally able to write neither the scalar form as a $3\times 3$ system nor such a system in scalar form.
\subsection{Scalar form}
\subsubsection{First order intertwiner}
In seeking to generalise the Laplace maps to the third order case we consider intertwining relations of the form
\[L^{\sigma_i}_{123}L_i=L^{\sigma_i}_iL_{123}.\]
For simplicity of notation we will consider $i=1$ and write $\sigma_1=\sigma.$

This definition of the first order intertwiner is a natural, formal generalization of the classical case. It would imply that if $\phi\in\ker L_{123}$ then $L_i\phi\in\ker L^{\sigma_i}_{123}.$

Apply the operators $\Theta_3,\Theta_2, \Theta_1$ and $\Theta_1^2$ to obtain the tower of equations
\begin{eqnarray}
L^\sigma_{123}L_1&=&L^\sigma_1L_{123}\label{0}\\
L^\sigma_{12}L_1&=&L^\sigma_1L_{12}\label{3}\\
L^\sigma_{13}L_1&=&L^\sigma_1L_{13}\label{2}\\
L^\sigma_{23}L_1+L^\sigma_{123}&=&L^\sigma_1L_{23}+L_{123}\label{1}\\
2L^\sigma_{23}&=&2L_{23}\label{11}
\end{eqnarray}

Equations (\ref{3}) and (\ref{2}) are Laplace maps of the earlier kind (with the label $1$ replacing $2$)  so that we immediately deduce:
\begin{eqnarray}
L^\sigma_2&=&L_2\nonumber\\
L^\sigma_3&=&L_3\nonumber\\
I^\sigma_{13}&=&I_{31}\nonumber\\
I^\sigma_{12}&=&I_{21}\nonumber\\
L^\sigma_1&=&I_{31}L_1I_{31}^{-1}=I_{21}L_1I_{21}^{-1}\label{con1}\\
I^\sigma_{21}-I_{21}&=&I_{21}-I_{12}+(\log I_{21}),_{12}\nonumber\\
I^\sigma_{31}-I_{31}&=&I_{31}-I_{13}+(\log I_{31}),_{13}.\nonumber
\end{eqnarray}

Equation (\ref{11}) is consistent with the above and requires in addition,
\[I^\sigma_{23}=I_{23},\quad I^\sigma_{32}=I_{32}.\]
Equation (\ref{con1}) on the other hand implies a constraint on the invariants, a condiation that requires satisfaction, if the Laplace map is to exist. The condition is
\[\left(\frac{I_{31}}{I_{21}}\right),_1=0.\]
 
If we restrict attention to the differential ring in the coefficients of the differential operators, i.e. we do not employ any extension, then, up to a multiplicative scalar, we conclude that
\[I_{21}=I_{31}.\] Unfortunately this relation is not preserved under the Laplace map:
\begin{eqnarray}
I^\sigma_{21}-I^\sigma_{31}&=&I^\sigma_{12}+[L^\sigma_1,L_2]-I^\sigma_{13}-[L^\sigma_1,L_3]\nonumber\\
&=&I_{21}-I_{31}+[L_1-(\log I_{21}),_1,L_2]-[L_1-(\log I_{31}),_1,L_3]\nonumber\\
&=&2I_{21}-I_{12}-2I_{31}+I_{13}+(\log I_{31}),_{13}-(\log I_{21}),_{12}\nonumber\\
&=&I_{13}-I_{12}+\left((\log I_{31}),_3-(\log I_{21}),_2\right),_1\nonumber\\
&\neq& 0\nonumber
\end{eqnarray}

We seek to understand if further constraints arise from equations (\ref{1}) and (\ref{0}).

Dealing with equation (\ref{1}) first, we can write it, using the expansion in invariants, as an expression relating $I^\sigma_{123}$ to $I_{321}:$
\begin{eqnarray}
L^\sigma_{123}-L^\sigma_1L^\sigma_{23}&=&L_{123}-L_{23}L_1\nonumber\\
L^\sigma_1L_2L_3+I_{21}L_3+I_{31}L_2&=&L_1L_2L_3+I_{12}L_3+I_{13}L_2\nonumber\\
+I_{23}L^\sigma_1+I^\sigma_{123}-L^\sigma_1(I_{23}+L_2L_3)&&+(L_{23}-L_2L_3)L_1+I_{123}\nonumber\\
I_{21}L_3+I_{31}L_2-I_{23},_1+I^\sigma_{123}&=&[L_1,L_2L_3]+I_{12}L_3+I_{13}L_2+I_{123}\nonumber\\
I_{31}L_2-I_{23},_1+I^\sigma_{123}&=&L_2(I_{31}-I_{13})+I_{13}L_2+I_{123}\nonumber\\
I^\sigma_{123}&=&I_{123}+I_{31},_2-I_{13},_2+I_{23},_1\nonumber\\
&=&I_{321}+I_{23},_1.\nonumber
\end{eqnarray}

Finally we deal with equation (\ref{0}). 

Once more using the invariant expansion of $L_{123}$ and $L^\sigma_{123}:$
\[(L^\sigma_1L^\sigma_2L^\sigma_3+I^\sigma_{23}L^\sigma_1+I^\sigma_{13}L^\sigma_2+I^\sigma_{12}L^\sigma_3+I^\sigma_{123})L_1\]
\[=L^\sigma_1(L_2L_3L_1+I_{23}L_1+I_{31}L_2+I_{21}L_3+I_{231})\]
Incorporating what we have learnt already,
\[(L^\sigma_1L_2L_3+I_{23}L^\sigma_1+I_{31}L_2+I_{21}L_3+I_{321}+I_{23},_1)L_1\]
\[=L^\sigma_1(L_2L_3L_1+I_{23}L_1+I_{31}L_2+I_{21}L_3+I_{231})\]
\[=L^\sigma_1L_2L_3L_1+L^\sigma_1I_{23}L_1+I_{31}L_1L_2+I_{21}L_1L_3+L^\sigma_1I_{231}\]
and we undertake the following manipulations
\begin{eqnarray}
I_{21}[L_3,L_1]+I_{31}[L_2,L_1]+(I_{321}+I_{23},_1)L_1&=&[L^\sigma_1,I_{23}]L_1+L^\sigma_1I_{231}\nonumber\\
I_{21}I_{13}-2I_{21}I_{31}+I_{31}I_{12}&=&L^\sigma_1I_{231}-I_{231}L_1\label{con2}
\end{eqnarray}

Summarising: the existence of a Laplace map of the form $L^\sigma_{123}L_1=L^\sigma_1L_{123}$ requires conditions on the form of the third order operator which can be expressed as the constraints on invariants arising from the three different expressions, (\ref{con1}) and (\ref{con2}), for $L_1^\sigma$ in terms of $L_1.$ These are succinctly written as the pair
\begin{eqnarray}
I_{31}I_{231},_1-I_{31},_1I_{231}&=&I_{31}\left(I_{21}I_{13}-2I_{21}I_{31}+I_{31}I_{12}\right)\nonumber\\
I_{21}I_{231},_1-I_{21},_1I_{231}&=&I_{21}\left(I_{21}I_{13}-2I_{21}I_{31}+I_{31}I_{12}\right)\nonumber
\end{eqnarray}
or, if the choice $I_{21}=I_{31}$ is made
\[\left(\frac{I_{231}}{I_{21}}\right),_1=I_{13}+I_{12}-2I_{21}.\]

These conditions amount to differential constraints on the coefficients $a_i,\,a_{ij}$ and $a_{ijk}$.

\subsubsection{Deformed first order intertwiner}
It is natural to ask if we can escape the constraints by considering the deformed relation:
\begin{eqnarray}
L^\sigma_{123}L'_1&=&L'^\sigma_1L_{123}\label{def0}\\
L^\sigma_{12}L'_1&=&L'^\sigma_1L_{12}\label{def3}\\
L^\sigma_{13}L'_1&=&L'^\sigma_1L_{13}\label{def2}\\
L^\sigma_{23}L'_1+L^\sigma_{123}&=&L'^\sigma_1L_{23}+L_{123}\label{def1}\\
L^\sigma_{23}&=&L_{23}\label{def11}
\end{eqnarray}

From (\ref{def3}) and (\ref{def2}) we will obtain 
\begin{eqnarray}
I^\sigma_{12}&=&I_{21}-\phi,_2\nonumber\\
\left(\frac{I^\sigma_{12}}{\phi}\right),_1&=&I^\sigma_{12}-I_{12}\nonumber\\
I^\sigma_{13}&=&I_{31}-\phi,_3\nonumber\\
\left(\frac{I^\sigma_{13}}{\phi}\right),_1&=&I^\sigma_{13}-I_{13}\nonumber.
\end{eqnarray}

Apparently we do not thereby escape constraints since the integrability conditions on $\phi$ require
\begin{eqnarray}
(I^\sigma_{12}-I_{21}),_3&=&(I^\sigma_{13}-I_{31}),_2\nonumber\\
\end{eqnarray}
and, writing 
$$\frac{I^\sigma_{12}}{I^\sigma_{13}}=\frac{I^\sigma_{12}}{\phi}\frac{\phi}{I^\sigma_{13}},$$
we get
\[\left(\frac{I^\sigma_{12}}{I^\sigma_{13}}\right),_1=\frac{I^\sigma_{12}I_{13}-I_{12}I^\sigma_{13}}{(I^\sigma_{13})^2}\phi\]
and hence
\[\phi=\frac{I^\sigma_{12},_1I_{13}-I_{12}I^\sigma_{13},_1}{I^\sigma_{12}I_{13}-I_{12}I^\sigma_{13}}.\]

As before there is a limit in which $\phi$ but not $\frac{\phi,_1}{\phi}$ tend to zero and this corrresponds to the previous relations: $I^\sigma_{12}=I_{21}$ etc.

Again, study of this situation is postponed to another time.

\subsubsection{Second order intertwiner}

A second possibility for implementing a type of Laplace transform is to consider a second order intertwiner,
\begin{equation}\label{second}
L^\sigma_{123}L_{12}=L^\sigma_{12}L_{123}
\end{equation}

We proceed as before and indeed it turns out that this imposes fewer constraints on the invariants of $L_{123}.$

$\Theta_3$ yields only trivial identities. The tower of identities obtained by applying $\Theta_1$ and $\Theta_2$ is:
\begin{eqnarray}
L^\sigma_{123}L_{12}&=&L^\sigma_{12}L_{123}\label{sec0}\\
L^\sigma_{23}L_{12}+L^\sigma_{123}L_2&=&L^\sigma_2L_{123}+L^\sigma_{12}L_{23}\label{sec1}\\
L^\sigma_{13}L_{12}+L^\sigma_{123}L_1&=&L^\sigma_1L_{123}+L^\sigma_{12}L_{13}\label{sec2}\\
L^\sigma_{23}L_2&=&L^\sigma_2L_{23}\label{sec11}\\
L^\sigma_{13}L_1&=&L^\sigma_1L_{13}\label{sec22}\\
L^\sigma_3L_{12}+L^\sigma_{23}L_1+L^\sigma_{13}L_2+L^\sigma_{123}&=&L_{123}+L^\sigma_2L_{13}+L^\sigma_1L_{23}+L^\sigma_{12}L_3\label{sec12}
\end{eqnarray}

Again equations (\ref{sec11}) and (\ref{sec22}) are of the classical form for a Laplace transformation and they give us
\begin{eqnarray}
L^\sigma_3&=&L_3\nonumber\\
I^\sigma_{23}&=&I_{32}\nonumber\\
I^\sigma_{13}&=&I_{31}\nonumber\\
L^\sigma_2&=&I_{32}L_2I_{32}^{-1}\nonumber\\
L^\sigma_1&=&I_{31}L_1I_{31}^{-1}.\nonumber
\end{eqnarray}

In particular, no constraint arises at this level.

Equation (\ref{sec12}) is dealt with using the invariant expansions
\[L^\sigma_{123}=L^\sigma_1L^\sigma_2L_3+I_{32}L^\sigma_1+I_{31}L^\sigma_2+I^\sigma_{12}L_3+I^\sigma_{123}\]
\[L_{123}=L_3L_2L_1+I_{32}L_1+I_{31}L_2+I_{21}L_3+I_{321}\]
and
\[L_{13}=L_3L_1+I_{31},\quad L_{23}=L_3L_2+I_{32},\quad L^\sigma=L^\sigma_1L^\sigma_2+I^\sigma_{12},\]
\[L^\sigma_{23}=L^\sigma_2L_3+I_{32},\quad L^\sigma_{13}=L^\sigma_1L_3+I_{31},\quad L_{12}=L_2L_1+I_{21}.\]

The result is an expression for $I^\sigma_{123}:$
\[I^\sigma_{123}=I_{321}+I_{32},_1-I_{21},_3+I_{31},_2.\]

We anticipate that equations (\ref{sec1}) and (\ref{sec2}) will give expressions for $I^\sigma_{12}$ and $I^\sigma_{21}$ and no constraints.

Write (\ref{sec1}) as
\[L^\sigma_{123}L_2-L^\sigma_{12}L_{23}=L_2^\sigma L_{123}-L^\sigma_{23}L_{12}.\]

Using the expansions
\[L^\sigma_{123}=L^\sigma_{12}L^\sigma_3+I^\sigma_{13}L^\sigma_2+I^\sigma_{23}L^\sigma_1+I^\sigma_{123}\]
\[L_{123}=L_3L_{12}+I_{32}L_1+I_{31}L_2+I_{321}-I_{21},_3\]
the relation simplifies to
\[(I^\sigma_{123}-I_{31},_2)L_2-L^\sigma_2(I_{321}-I_{21},_3)=L^\sigma_{12}I_{32}-I_{32}L_{12}+L^\sigma_2I_{32}L_1-I_{32}L^\sigma_1L_2.\]

Then
\begin{eqnarray}
(I^\sigma_{123}-I_{31},_2)L_2-L^\sigma_2(I_{321}-I_{21},_3)&=&L^\sigma_{12}I_{32}-I_{32}L_{12}+I_{32}L_2L_1-I_{32}L^\sigma_1L_2\nonumber\\
&=&(L^\sigma_1L^\sigma_2+I^\sigma_{12})I_{32}-I_{32}I_{21}-I_{32}L^\sigma_1L_2\nonumber\\
&=&I_{32},_1L_2+(I^\sigma_{12}-I_{21})I_{32}.\nonumber
\end{eqnarray}

Define
\[J_{321}=I_{321}-I_{21},_3=I^\sigma_{123}-I_{31},_2-I_{32},_1.\]

Then
\begin{equation}\label{J1}
L^\sigma_2J_{321}-J_{321}L_2=(I_{21}-I^\sigma_{12})I_{32}
\end{equation}
and from equation (\ref{sec2}), by transposition of $1$ and $2$,
\begin{equation}\label{J2}
L^\sigma_1J_{312}-J_{312}L_1=(I_{12}-I^\sigma_{21})I_{31}.
\end{equation}

Note that, following from the properties of $I_{ijk}$  under transposition
\[J_{321}=J_{312}.\]

Using the relations $L^\sigma_2I_{32}=I_{32}L_2$ and $L^\sigma_1I_{31}=I_{31}L_1$ we get expressions for $I^\sigma_{21}$ and $I^\sigma_{12}:$
\begin{eqnarray}
I^\sigma_{12}&=&I_{21}-\left(\frac{J_{321}}{I_{32}}\right),_2\nonumber\\
I^\sigma_{21}&=&I_{12}-\left(\frac{J_{312}}{I_{31}}\right),_1.\nonumber
\end{eqnarray}

Finally we must analyse the relation (\ref{sec0}) itself:
\[L^\sigma_{123}L_{12}=L^\sigma_{12}L_{123}.\]

At this point all the $I^\sigma$ invariants are defined in terms of the $I$ invariants as are the $L^\sigma_i$ so the most we can hope for here is a consistent set of relations but we might expect, instead, constraints to arise.

We use the same invariant expansions as in the previous calculations to arrive at,
\[(I_{31}L^\sigma_2+I_{32}L^\sigma_1+I^\sigma_{123})L_{12}=L^\sigma_{12}(I_{32}L_1+I_{31}L_2+J_{321}).\]

We can use the $J_{321}$ relations (\ref{J1}) and  (\ref{J2}) to move the $L^\sigma_1$ and $L^\sigma_2$ operators from left to right to obtain
\begin{eqnarray}
I^\sigma_{123}L_{12}&=&I_{32},_1L_2L_1-I_{32}L^\sigma_1+I^\sigma_{12}I_{32}L_1\nonumber\\
&&+I_{31},_2L_1L_2-I_{31}L^\sigma_2I_{12}+I^\sigma_{21}I_{31}L_2\nonumber\\
&&+J_{321}L_1L_2+I_{31}(I_{12}-I^\sigma_{21})L_2+I^\sigma_{12}J_{321}\nonumber\\
&&+L^\sigma_1(I_{21}-I^\sigma_{12})I_{32}\nonumber
\end{eqnarray}
 Rearranging we obtain the relations between invariants
\begin{eqnarray}\label{C}
J_{321}I_{12}&=&(I^\sigma_{12}I_{32})L_1-L^\sigma_1(I^\sigma_{12}I_{32})\nonumber\\
&&+(I_{31}I_{12})L_2-L^\sigma_2(I_{31}I_{12})\nonumber.
\end{eqnarray}

Since all the objects with $\sigma$ superfix are already defined, this relation constitutes a constraint on the invariants. Again this amounts to a differential constraint on the coefficients $a_i,\,a_{ij}$ and $a_{ijk}$.

\subsection{System form}
In \cite{A} Laplace type maps are developed for third order systems of the form
\[
\left(
\begin{array}{ccc}
L_1 & h_{12} & h_{13}\\
h_{21} & L_2 & h_{23}\\
h_{31} & h_{32} & L_3
\end{array}
\right)
\left(
\begin{array}{c}
\phi_1\\
\phi_2\\
\phi_3
\end{array}
\right)
=
\left(
\begin{array}{c}
0\\
0\\
0
\end{array}
\right)
\]
where $L_i=\partial_i+h_{ii}.$ In this section we provide a slight generalization of those results.

The set of transformations under which invariants are defined are taken to be conjugations of the matrix differential operator by diagonal matrices of three arbitrary functions
\[G=
\left(
\begin{array}{ccc}
g_1 & 0 & 0\\
0 &g_2 & 0\\
0 & 0 & g_3
\end{array}
\right)
\]
and the invariants are
\begin{eqnarray}
(12)&=&h_{12}h_{21}\nonumber\\
(23)&=&h_{23}h_{32}\nonumber\\
(31)&=&h_{31}h_{13}\nonumber\\
(123)&=&h_{12}h_{23}h_{31}\nonumber\\
(132)&=&h_{13}h_{32}h_{21}\nonumber\\
{[}12]&=&h_{11},_2-h_{22},_1+\frac12(\log\frac{h_{12}}{h_{21}}),_{12}\nonumber\\
{[}13]&=&h_{11},_3-h_{33},_1+\frac12(\log\frac{h_{13}}{h_{31}}),_{13}\nonumber\\
{[}23]&=&h_{22},_3-h_{33},_2+\frac12(\log\frac{h_{23}}{h_{32}}),_{23}\nonumber.
\end{eqnarray}

The $(ij)$ and $(ijk)$ are totally symmetric. The $[ij]$ are antisymmetric.

Note that this is not an independent set. In fact
\[(12)(23)(31)=(123)(132),\]
and
\[[12],_3+[23],_1+[31],_2=\frac12(\log\frac{(123)}{(132)}),_{123}.\]

A number of variant canonical forms are possible. For example, choosing $g,_1+h_{11}g=0$ and $g_1=g,_1,\,g_2=gh_{21},\,g_3=gh_{21}h_{32}$ we obtain
\begin{equation}\label{can}
G^{-1}\left(
\begin{array}{ccc}
L_1 & h_{12} & h_{13}\\
h_{21} & L_2 & h_{23}\\
h_{31} & h_{32} & L_3
\end{array}
\right)G
=
\left(
\begin{array}{ccc}
\partial_1 & (12) & (132)\\
1 & \partial_2+I_2 & (23)\\
\frac{(13)}{(132)} & 1 & \partial_3+I_3
\end{array}
\right)
\end{equation}
where 
\[I_2,_1=[21]+\frac12(\log (12)),_{12},\] and
\[I_3,_1=[31]+(\log (132))_{13}-\frac12(\log (13)),_{13}.\]

The {\em deformed} intertwining relation generalises that for the $2\times 2$ case. For example,
\[
\left(
\begin{array}{ccc}
L^\sigma_1 & h^\sigma_{12} & h^\sigma_{13}\\
h^\sigma_{21} & L^\sigma_2 & h^\sigma_{23}\\
h^\sigma_{31} & h^\sigma_{32} & L^\sigma_3
\end{array}
\right)
\left(
\begin{array}{ccc}
\partial_1+u_{11} & h_{12}-h^\sigma_{12} & h_{13}-h^\sigma_{13}\\
0 & \partial_1+u_{22} & 0\\
0 & 0 & \partial_1+u_{33}
\end{array}
\right)
\]
\[=
\left(
\begin{array}{ccc}
\partial_1+l_{11} & 0 & 0\\
h^\sigma_{21}-h_{21} & \partial_1+l_{22} &0\\
h^\sigma_{31}-h_{31} & 0 & \partial_1+l_{33}
\end{array}
\right)
\left(
\begin{array}{ccc}
L_1& h_{12} & h_{13}\\
h_{21} & L_2 & h_{23}\\
h_{31} & h_{32} & L_3
\end{array}
\right)
\]
from which follow
\begin{eqnarray}
l_{22}&=&u_{22}\nonumber\\
l_{33}&=&u_{33}\nonumber\\
L^\sigma_2&=&L_2\nonumber\\
L^\sigma_3&=&L_3\nonumber\\
h^\sigma_{23}&=&h_{23}\nonumber\\
h^\sigma_{32}&=&h_{32}\nonumber
\end{eqnarray}
and
\[u_{22}=h_{11}-(\log h_{21}),_1+(12)^\sigma\alpha=h^\sigma_{11}+(\log h^\sigma_{12}),_1+(12)\alpha,\]
\[u_{33}=h_{11}-(\log h_{31}),_1+(13)^\sigma\beta=h^\sigma_{11}+(\log h^\sigma_{13}),_1+(13)\beta.\]

Here $\alpha$ and $\beta$ are functions of invariants satisfying,
\[\frac{\alpha}{\beta}=\frac{(132)^\sigma(13)}{(132)(12)^\sigma}=\frac{(123)(13)^\sigma}{(123)^\sigma(12)}\]
in terms of which the Laplace map takes the form 
\begin{eqnarray}
(12)^\sigma-(12)&=&[12]-\frac12(\log(12)),_{12}+\left((12)^\sigma\alpha\right),_2\nonumber\\
(13)^\sigma-(13)&=&[13]-\frac12(\log (13)),_{13}+\left((13)^\sigma\beta\right),_3\nonumber\\
(23)^\sigma-(23)&=&0\nonumber\\
{[}12]^\sigma-[12]&=&-\frac12\left(\log((12)(12)^\sigma\alpha^2)\right),_{12}\nonumber\\
{[}13]^\sigma-[13]&=&-\frac12\left(\log((13)(13)^\sigma\beta^2)\right),_{13}\nonumber\\
{[}23]^\sigma-[23]&=&0\nonumber\\
(132)^\sigma-(132)&=&(23)\left(\log\frac{(12)}{(123)}\right),_1-\alpha(12)^\sigma+\beta(13)^\sigma\nonumber\\
(123)^\sigma-(123)&=&(23)\left(\log\frac{(13)}{(132)}\right),_1+\alpha(12)^\sigma-\beta(13)^\sigma.\nonumber
\end{eqnarray}

The Laplace maps respect the differential identities satisfied by the antisymmetric invariants.
\begin{eqnarray}
{[}12]^\sigma,_3+[23]^\sigma,_1+[31]^\sigma,_2&=&[12],_3+[23],_1+[31],_2+\frac12\left(\log\frac{(13)(13)^\sigma\beta^2}{(12)(12)^\sigma\alpha^2}\right),_{123}\nonumber\\
&=&\frac12\left(\log\frac{(123)(13)(13)^\sigma\beta^2}{(132)(12)(12)^\sigma\alpha^2}\right),_{123}\nonumber\\
&=&\frac12\left(\log\frac{(123)^\sigma}{(132)^\sigma}\right),_{123}.
\end{eqnarray}

The other algebraic constraint on the symmetric invariants 
\[(123)^\sigma(132)^\sigma=(12)^\sigma(23)^\sigma(31)^\sigma\]
provides a second (complicated) condition, not identically satisfied, on $\alpha$ and $\beta,$ thus determining them.

Note that in \cite{A} the simplifying choice $u_{22}=u_{33}$ has been made which leads to the relation $\alpha=\beta.$ It was not appreciated at that time that such a choice leads to constraints on the invariants. As is seen above allowing $u_{22}\neq u_{33}$ leads to more complexity. This is currently being explored.

\subsection{Scalarizability}

Because the differential operator of the $3\times 3$ system is defined over a non-commutative ring one cannot, in general, reduce to a single scalar equation as might over a field. But it is clear that the vanishing of any of the $h_{ij}$ is a sufficient condition to do so and we say that in such a case the system is {\em scalarizable}. 

For example, if $h_{23}=0$ then we can write down a scalar equation for $\phi_2$ of the form
\[(\tilde L_3\tilde L_1\tilde L_2-\tilde L_3(12)-(13)\tilde L_2+(132))\tilde \phi_2=0,\]
where
\begin{eqnarray}
\tilde L_1&=&L_1\nonumber\\
\tilde L_2&=&h^{-1}_{21}L_2h_{21}\nonumber\\
\tilde L_3&=&h_{13}L_3h^{-1}_{13}\nonumber\\
\tilde \phi_2&=&h_{21}^{-1}\phi_2\nonumber.
\end{eqnarray}

By comparing with the invariant expansion
\[\tilde L_3\tilde L_1\tilde L_2+\tilde I_{32}\tilde L_1+\tilde I_{31}\tilde L_2+\tilde I_{12}L_3+\tilde I_{312}\]
the corresponding invariants $I_{ij}$ and $I_{123}$ are seen to be
\begin{eqnarray}
\tilde I_{32}&=&0\nonumber\\
\tilde I_{23}&=&[23]+(\log(132)),_{23}\nonumber\\
\tilde I_{31}&=&-(13)\nonumber\\
\tilde I_{13}&=&-(13)+[13]+\frac12(\log(13)),_{13}\nonumber\\
\tilde I_{12}&=&-(12)\nonumber\\
\tilde I_{21}&=&-(12)+[21]+\frac12(\log(12)),_{12}\nonumber\\
\tilde I_{312}&=&(132)-(12),_3\nonumber
\end{eqnarray}

using $(23)=0.$

Finally we would like to see that the system Laplace map corresponds with a scalar one. It suffices to check that the invariants transform in the same way. The restriction $(23)=0$ is consistent with (but not equivalent to) the choice $\alpha=\beta=0$ which follows from the assumption $u_{11}=h_{11}$ and $l_{11}=h^\sigma_{11},$ corresponding to the undeformed $3\times 3$ system. This in turn implies
\begin{eqnarray}
\partial_1+u_{22}&=&h_{21}L_1h^{-1}_{21}\nonumber\\
&=&{(h^\sigma_{12})}^{-1}L^\sigma_1h^\sigma_{12}\nonumber\\
\partial_1+u_{33}&=&h_{31}L_1h^{-1}_{31}\nonumber\\
&=&{(h^\sigma_{13})}^{-1}L^\sigma_1h^\sigma_{13}\nonumber\\
\end{eqnarray}
and consequently
\[(h^\sigma_{12}h_{21})L_1(h^\sigma_{12}h_{21})^{-1}=(h^\sigma_{13}h_{31})L_1(h^\sigma_{13}h_{31})^{-1},\]
or
\[h^\sigma_{12}h_{21}=h^\sigma_{13}h_{31},\]
or, in invariant terms,
\[\frac{(123)^\sigma}{(13)^\sigma}=\frac{(123)}{(12)}.\]

 We impose the limit $\alpha,\beta\rightarrow 0$ and $\frac{\alpha}{\beta}=1$ on what follows.

Consider the $\tilde I^\sigma_{13}$ and $\tilde I^\sigma_{31}$ invariants arising from the first order scalar Laplace map. From the results on the first order intertwiners,
\begin{eqnarray}
\tilde I^\sigma_{13}&=&\tilde I_{31}=-(13)\nonumber\\
\tilde I^\sigma_{31}&=&2\tilde I_{31}-\tilde I_{13}+(\log\tilde I_{31}),_{13}\nonumber\\
-(13)^\sigma&=&-2(13)+(13)-[13]-\frac12(\log(13)),_{13}+(\log(13)),_{13}\nonumber\\
&=&-(13)-[13]+\frac12(\log(13)),_{13}\nonumber.
\end{eqnarray}
which rearranges to
\[(13)^\sigma-(13)=[13]-\frac12(\log(13)),_{13}\]
as expected.

For the antisymmetric invariant we have,
\begin{eqnarray}
-(13)^\sigma+[13]^\sigma+\frac12(\log(13)^\sigma)_{13}&=&-(13)\nonumber\\
{[}13]^\sigma&=&(13)^\sigma-(13)-\frac12(\log (13)^\sigma)_{13}\nonumber\\
{[}13]^\sigma-[13]&=&-\frac12(\log (13)^\sigma(13))_{13}\nonumber
\end{eqnarray}

The $I_{12}$ relation follows simply from:
\begin{eqnarray}
\tilde I^\sigma_{12}&=&\tilde I_{21}\nonumber\\
- (12)^\sigma&=&-(12)+[21]+\frac12(\log(12)),_{12}\nonumber\\
(12)^\sigma-(12)&=&[12]-\frac12(\log(12)),_{12}\nonumber
\end{eqnarray}
as expected and so on.

The condition that $h_{23}=0$ implies $(23)=0$ and for the system Laplace map this gives
\[(132)^\sigma=(132)\]
which we wsih to show is equivalent to 
\[I^\sigma_{123}=I_{321}+I_{23},_1,\]
the Laplace map result for the scalar operator. Rewriting the system relation using the permutation relations on three index objects,
\begin{eqnarray}
\tilde I^\sigma_{312}-\tilde I^\sigma_{12},_3&=&\tilde I_{312}-\tilde I_{12},_3\nonumber\\
\tilde I^\sigma_{132}-\tilde I^\sigma_{12},_3&=&\tilde I_{312}-\tilde I_{12},_3\nonumber\\
\tilde I^\sigma_{123}+\tilde I^\sigma_{32},_1-\tilde I^\sigma_{23},_1-\tilde I^\sigma_{12},_3&=&\tilde I_{321}+\tilde I_{12},_3-\tilde I_{21},_3-\tilde I_{12},_3\nonumber\\
\tilde I^\sigma_{123}+\tilde I_{32},_1-\tilde I_{23},_1-\tilde I_{21},_3&=&\tilde I_{321}-\tilde I_{21},_3\nonumber\\
\tilde I^\sigma_{123}&=&\tilde I_{321}+\tilde I_{23},_1+\tilde I_{32},_1\nonumber\\
&=&\tilde I_{321}+\tilde I_{23},_1\nonumber
\end{eqnarray}
as desired, since $\tilde I_{32}=0.$

\subsubsection{Conclusion}
This paper has examined in detail a natural generalization of the classical Laplace map on invariants of second order, partial differential operators to third order operators of both scalar and system forms and compared them in the overlap where a system is scalarizable. The basic obervation is that such intertwining maps of scalar operators can only exist where prior conditions exist on the invariants. It is shown by example that these can be compatible with the scalarizability of a system for which more general (unconstrained) Laplace maps exist.

There is a further generalization of the notion of Laplace map which would allow us to avoid the constraints in the case of the second order intertwining relation. We call this a {\em weak Laplace map}: Consider the inhomogeneous partial differential equation
\[L_{123}\phi=\psi\]
and suppose that the intertwining relation holds only up to an unspecified function, $K$:
\[L^\sigma_{12}L_{123}-L^\sigma_{123}L_{12}=K.\]
Then the conditions (\ref{sec1}) to (\ref{sec12}) hold but not (\ref{sec0}) and the relations between the $I^\sigma$ and the $I$ are as in the last section but without the constraint (\ref{C}). This allows us to define the weak Laplace map:
\begin{eqnarray}
L^\sigma_{12}L_{123}\phi&=&L^\sigma_{12}\psi\nonumber\\
L^\sigma_{123}L_{12}\phi&=&L^\sigma_{12}\psi-K\phi\nonumber\\
L^\sigma_{123}\phi^\sigma&=&\psi^\sigma\nonumber
\end{eqnarray}
where,
\[
\phi\mapsto \phi^\sigma =L_{12}\phi,\quad \psi\mapsto\psi^\sigma=L^\sigma_{12}\psi-K\phi.
\]

There seems to be some scope here for further work.

Apart from the study of weak maps and the deformed intertwining relations, other natural questions include:
\begin{itemize}
\item What are the general constraints on the invariants of a $3\times 3$ (or higher order) system that make it scalarizable?
\item What conditions need to be satisfied by the invariants of a scalar operator in order for it to be written in system form?
\item Does the second order intertwining map also correspond to a Laplace map of a $3\times 3$ system?
\end{itemize}

\section{Acknowledgements}

I would like to thank the two anonymous referees for their thorough reading of the manuscript and their suggestions for clarification and improvement.

This paper was written during the last weeks and days of my close friend and colleague, Jon Nimmo. It is appropriate to record here my appreciation of Jon's support, generous encouragement and humour over many years in Glasgow.

\end{document}